# Giant Second-Harmonic Generation in 3R-$MoS_2$/MLM Hybrid Metasurfaces Cavities

Omar A. M. Abdelraouf [a,*]

[a] Institute of Materials Research and Engineering, Agency for Science, Technology, and Research (A*STAR), 2 Fusionopolis Way, #08-03, Innovis, Singapore 138634, Singapore.

## ABSTRACT

Nonlinear 2D materials such as 3R-phase molybdenum disulfide (3R-$MoS_2$) offer strong second-order optical nonlinearities in an atomically thin platform, making them attractive for on-chip frequency conversion, quantum light generation, and integrated nonlinear nanophotonics. However, the second harmonic generation (SHG) efficiency of monolayer or few-layer 3R-$MoS_2$ deposited on planar substrates remains fundamentally limited by weak light–matter interaction, poor phase matching, and small interaction volumes. Multi-layer metasurfaces (MLMs) provide expanded geometric and material degrees of freedom, enabling dual-resonant cavities that can independently and simultaneously enhance the fundamental (pump) and second-harmonic fields at the 3R-$MoS_2$ location. Yet, traditional trial-and-error or forward-only optimization becomes impractical for designing such dual-resonant, multi-parameter structures due to the vast design space and the need to co-optimize linear and nonlinear responses. Here, we introduce NanoPhotoNet-PINL, a physics informed AI-driven inverse design framework based on a hybrid one-dimensional convolutional neural network and deep neural network autoencoder, tailored for nonlinear MLMs metasurfaces. The model directly maps target dual-resonant reflection spectra at the fundamental and second-harmonic wavelengths to the required multi-layer geometries and material compositions that maximize the effective nonlinear overlap with an embedded 3R-$MoS_2$ sheet. By integrating Maxwell-based nonlinear electrodynamics into the inverse design loop, we compute the second-harmonic conversion efficiency and modal overlap factors for each predicted MLMs design, enabling physics-guided training and evaluation. Our approach achieves an inverse-design prediction efficiency exceeding ~99.2% along the linear spectral manifold, while the optimized dual-resonant MLMs yield more than three orders of magnitude enhancement in SHG intensity compared to a bare 3R-$MoS_2$ flake on a planar substrate. These substantial increases in nonlinear conversion and radiation efficiency position the 3R-$MoS_2$/MLMs platform as a scalable route towards compact, CMOS-compatible nonlinear metasurface devices for on-chip frequency converters, quantum light sources, and all-optical signal processing. Consequently, NanoPhotoNet-PINL establishes a generalizable paradigm for intelligent inverse design of nonlinear multi-layer metasurfaces and phase-matched dual-resonant cavities for high-efficiency second-order processes.

* Corresponding author. Email address: Omar_Abdelrahman@a-star.edu.sg

## 1. INTRODUCTION

Second-order nonlinear optical processes such as second harmonic generation (SHG) underpin a broad spectrum of emerging nanophotonic technologies, from on-chip coherent light sources and frequency converters to quantum light generation and ultrafast all-optical signal processing.[1-3] Atomically thin transition metal dichalcogenides (TMDs) have recently emerged as highly promising nonlinear media, owing to their large second-order susceptibilities, excitonic resonances in the visible and near-infrared, and compatibility with planar integration.[4-6] In particular, the 3R polytype of molybdenum disulfide (3R-$MoS_2$) exhibits broken inversion symmetry even in few-layer configurations, leading to strong bulk-like second-order responses that far exceed those of conventional centrosymmetric 2H phases when averaged over multiple layers.[7] While these intrinsic nonlinear properties are favorable, the transition of 3R-$MoS_2$ from nano-scale test structures to practical integrated nonlinear photonic circuits is limited by a critical bottleneck: the extremely small interaction volume and weak overlap between the guided or radiated optical modes and the atomically thin nonlinear layer on a planar substrate.[8] As a result, the SHG efficiency of bare 3R-$MoS_2$ films is typically low, especially under moderate pump intensities compatible with on-chip operation.

To overcome these limitations, it is imperative to engineer the local photonic environment surrounding the 3R-$MoS_2$ to simultaneously enhance the fundamental field, increase the second-harmonic field buildup, and tailor the radiation pattern for efficient out-coupling.[9-11] Metasurfaces, composed of subwavelength resonant nanostructures, provide a mature platform for manipulating light absorption,[12-27] amplitude, phase, and polarization of light at will, enabling strong field confinement and enhanced nonlinear interactions.[28-39] However, single-layer metasurfaces usually offer a limited set of geometric degrees of freedom, which constrains the ability to realize independently tunable dual resonances at the pump and SH wavelengths with optimal spatial overlap at the 3R-$MoS_2$ plane. Moreover, achieving high quality-factor resonances for both frequencies while maintaining fabrication tolerance and robust nonlinear response demands a nontrivial compromise between material losses, dispersion, and radiative coupling, which is often difficult to reach with simple planar designs.[40]

Multi-layer metasurfaces (MLMs), or cascaded metasurface architectures, markedly expand the design space by stacking multiple patterned layers and materials along the propagation direction.[41, 42] This vertical degree of freedom introduces a large number of independent parameters such as layer thickness, refractive index, nano-pillar width, and interlayer spacing that can be exploited to engineer high-Q, dual-resonant cavities where the fundamental and second-harmonic modes overlap spatially and spectrally with the 3R-$MoS_2$ sheet. In addition, the multi-layer configuration allows controlled breaking of $z$-symmetry, which is essential for optimizing far-field radiation of the second harmonic, suppressing parasitic substrate leakage, and tailoring the polarization response to match the nonlinear tensor elements of 3R-$MoS_2$.[43] Achieving such precise modal engineering is particularly important for nonlinear metasurfaces operating near excitonic resonances, where even small changes in field localization and phase profile can dramatically modify the SHG yield.

Despite their superior potential, the high dimensionality and complexity of MLMs make conventional design methodologies inefficient and often impractical. Trial-and-error approaches based on sweeping a small subset of parameters cannot capture the intricate interplay between the fundamental and second-harmonic resonances, the spatial mode profiles, and the nonlinear overlap integrals.[44] Forward-only optimization methods such as Genetic Algorithms (GAs) or Particle Swarm Optimization (PSO) can, in principle, explore larger design spaces, but they suffer from slow convergence and require thousands of full-wave electromagnetic simulations (e.g., Finite-Difference Time-Domain (FDTD) or Finite Element Method (FEM)) to approach sub-optimal solutions.[45-47] Furthermore, these methods struggle with the nonlinear inverse-design task of mapping a desired dual-resonant linear spectrum such as specified quality factors, resonance frequencies, and polarization properties to a physically realizable multilayer geometry that also maximizes the nonlinear SHG output. This difficulty is compounded by the "one-to-many" nature of inverse scattering, where multiple distinct MLMs configurations yield similar linear spectra yet exhibit different nonlinear performances due to variations in field distributions at the 3R-$MoS_2$ plane.

These challenges motivate the development of data-driven, physics-informed inverse-design frameworks that can efficiently synthesize high-dimensional MLMs geometries from target spectral and nonlinear performance requirements. Deep learning models, particularly convolutional neural networks (CNNs) and autoencoders, have shown promising capabilities in learning complex structure–response mappings in electromagnetics and nanophotonics, enabling rapid inverse design far beyond the reach of traditional numerical sweeps.[48-50] However, most previous efforts have focused on linear metasurfaces, simple single-layer configurations, or narrow frequency ranges, and often lack explicit incorporation of nonlinear physics. There remains a clear need for a generalized inverse-design framework that can jointly address the linear dual-resonant responses and the nonlinear SHG efficiency in multi-layer, multi-material metasurfaces incorporating 2D nonlinear media.[51, 52]

To address this gap, we develop NanoPhotoNet-PINL, a discriminative deep-learning framework that directly maps target dual-resonant spectral responses at the fundamental and SHG wavelengths to the multilayer geometry of a 3R-$MoS_2$-based MLMs cavity. The architecture integrates a 1D CNN encoder for spectral feature extraction, an adaptive average pooling layer to produce a compact latent representation, and a deep neural network decoder to reconstruct 12 geometric and material parameters defining the MLMs stack. Nonlinear electrodynamics, including the SHG source terms and overlap integrals, is embedded into the data-generation and evaluation pipeline so that each inverse-designed MLMs can be rigorously assessed in terms of SHG enhancement compared to a planar 3R-$MoS_2$ thin-film reference. The resulting framework provides a fast, reliable, and physics-guided route to engineer dual-resonant MLMs that realize over three orders of magnitude enhancement of SHG in 3R-$MoS_2$, opening new routes towards integrated nonlinear metasurface devices.

## 2. Methods

To enable robust inverse design of dual-resonant nonlinear MLMs integrating 3R-$MoS_2$, we first generate a comprehensive dataset spanning diverse geometric and material configurations that are relevant for simultaneous control of fundamental and SHG resonances. Each MLMs design comprises 1 to 5 vertically stacked layers with square-shaped nanopillars within a square unit cell. The use of a square unit cell imposes in-plane symmetry and simplifies numerical modeling, while multilayer stacking introduces crucial vertical degrees of freedom for tailoring the field profiles at the 3R-$MoS_2$ plane. High-index dielectrics and low-loss materials are selected to support high-Q resonances in the visible and near-infrared range relevant for typical pump–SHG wavelength pairs associated with 3R-$MoS_2$ excitonic transitions.[53] Geometric parameters such as layer heights ($h_i$), nanopillar refractive index ($n_i$), width ($w$), and lattice period ($P$) are sampled over broad ranges to cover weakly and strongly resonant regimes. For each configuration, we perform FDTD simulations under $x$-polarized plane-wave excitation with periodic boundary conditions in-plane and Perfectly Matched Layers (PMLs) along the propagation direction as shown in Figure 1a. The linear reflection spectra are recorded over a wavelength range from 350 nm to 900 nm, discretized into 1000 spectral points. Each simulation also includes a nonlinear post-processing step in which the fundamental field distribution *E(ω,r)* at the pump wavelength and the cavity mode at the second harmonic *E(2ω,r)* are used to evaluate the SHG source and radiation patterns for a 3R-$MoS_2$ sheet located at a designated interface within the MLMs stack. The nonlinear polarization density associated with SHG in 3R-$MoS_2$ is expressed as

$$P_i(2\omega) = \varepsilon_0 \sum_{j,k} \chi_{ijk}^{(2)} E_j(\omega) E_k(\omega) \tag{1}$$

where $\varepsilon_0$ is the vacuum permittivity, $\chi_{ijk}^{(2)}$ denotes the effective second-order nonlinear susceptibility tensor components of 3R-$MoS_2$, and $E_j(\omega)$ is the electric field component at the fundamental frequency $(\omega)$.[54] For a given polarization configuration and crystal orientation, this expression can be simplified to an effective scalar relation

$$P_{\mathrm{NL}}(2\omega) = \varepsilon_0 \chi_{\mathrm{eff}}^{(2)} E_{\parallel}^2(\omega) \tag{2}$$

where $E_{\parallel}(\omega)$ denotes the in-plane electric field component parallel to the dominant nonlinear tensor elements. The generated SHG field *E(2ω)* obeys the inhomogeneous Helmholtz equation

$$\nabla \times \nabla \times E(2\omega, r) - \frac{(2\omega)^2}{c^2} \varepsilon(2\omega, r) E(2\omega, r) = i(2\omega)\mu_0 P_{\mathrm{NL}}(2\omega, r) \tag{3}$$

where $\varepsilon(2\omega, r)$ is the permittivity distribution at the second harmonic frequency and $\mu_0$ is the vacuum permeability.[55] We solve this equation numerically within the FDTD framework, assuming that the SHG power remains sufficiently small such that pump depletion is negligible. The SHG conversion efficiency $\eta_{\mathrm{SHG}}$ is evaluated as

$$\eta_{\mathrm{SHG}} = \frac{P_{2\omega}}{P_{\omega}^2} = \frac{\int_{S_{\mathrm{out}}} \langle S_{2\omega} \cdot \hat{n} \rangle dS}{\left( \int_{S_{\mathrm{in}}} \langle S_{\omega} \cdot \hat{n} \rangle dS \right)^2} \tag{4}$$

where $P_{2\omega}$ is the total radiated SHG power through an output surface $S_{\mathrm{out}}$, $P_{\omega}$ is the pump power through an input surface $S_{\mathrm{in}}$, $\hat{n}$ is the outward normal, and $S_{\omega}$ and $S_{2\omega}$ are the time-averaged Poynting vectors at the pump and second harmonic frequencies, respectively. To quantify the enhancement provided by the MLMs cavity relative to a planar reference, we define the SHG enhancement factor (*EEF*) or amplification factor ($A_{SHG}$).

$$EF = A_{\mathrm{SHG}} = \frac{\eta_{\mathrm{SHG}}^{\mathrm{MLM}}}{\eta_{\mathrm{SHG}}^{\mathrm{TF}}} \tag{5}$$

where $\eta_{\mathrm{SHG}}^{\mathrm{MLM}}$ is the SHG efficiency with 3R-$MoS_2$ integrated into the MLMs cavity, and $\eta_{\mathrm{SHG}}^{\mathrm{TF}}$ is the efficiency of a 3R-$MoS_2$ thin film on an unstructured planar substrate under identical pump conditions. The modal field enhancement of SHG ($\Lambda$) in resonant cavities is closely related to the overlap of the fundamental and second-harmonic modes and the effective nonlinear modal volume. The standard overlap integral for a second-order process can be written as

$$\Lambda = \frac{\left|\int_V \chi_{\mathrm{eff}}^{(2)} E_{\parallel}^2(\omega,r) E_{\parallel}^*(2\omega,r)\, dV\right|}{\left(\int_V \varepsilon(\omega,r)|E(\omega,r)|^2 dV\right)\sqrt{\int_V \varepsilon(2\omega,r)|E(2\omega,r)|^2 dV}} \tag{6}$$

where the integrals extend over the nonlinear region containing 3R-$MoS_2$ and nearby field localization volume.[56] In high-Q cavities, the SHG efficiency can be approximated as scaling with the square of the fundamental quality factor $Q_{\omega}$, the second harmonic quality factor $Q_{2\omega}$, and the squared overlap $|\Lambda|^2$:

$$\eta_{\mathrm{SHG}} \propto Q_{\omega}^2 Q_{2\omega} |\Lambda|^2 \tag{7}$$

for fixed input intensity and material properties. This scaling relation emphasizes the necessity of dual-resonant designs where both frequencies are simultaneously confined and spatially overlapped at the 3R-$MoS_2$ sheet. The linear portion of the dataset that used to train the inverse model consists of pairs of reflection spectra (input) and MLMs geometries (output). Each reflection spectrum is represented as a 1000-point vector covering the same wavelength range as above, and captures both the fundamental and second harmonic resonances through their linear spectral signatures. The output vector encodes 12 geometric parameters, including the refractive indices, layer heights, nanopillar widths, and the lattice period. All scalar inputs and outputs are normalized to the range using min–max normalization. A total of 10,080 samples generated and partitioned into training (70%), validation (15%), and testing (15%) sets.

The NanoPhotoNet-PINL, AI-inverse framework is formulated as a one-dimensional convolutional autoencoder, consisting of a 1D CNN encoder, a global average pooling layer, a dense latent bottleneck, and a deep neural network decoder. The 1D CNN encoder processes the discretized reflection spectrum and extracts spectral features associated with resonance positions, linewidths, and background continuum. The fundamental operation of each convolutional layer is

$$y[n] = (x * w)[n] = \sum_{k=1}^{K} x[n-k+1] \cdot w[k] + b \tag{8}$$

where $x$ is the input spectrum or feature map, $w$ represents learnable kernel weights, $K$ is the kernel size, $y$ is the output feature map, and $b$ is the bias term. Multiple convolutional layers with increasing channel depth and strided convolutions are stacked to hierarchically learn local and global spectral features, including sharp, high-Q resonances and broader low-Q modes. Following the convolutional layers, an adaptive global average pooling (GAP) 1D layer condenses the variable-length feature maps into a fixed-size vector of length 256, independent of the exact spectral sampling density. The GAP operation can be expressed as

$$y[c] = \frac{1}{T} \sum_{t=0}^{T-1} x[c,t] \tag{9}$$

where $c$ indexes the channel dimension, $T$ is the temporal (spectral) length, and $x[c,t]$ denotes the feature activation at channel $c$ and position $t$. This pooling step promotes translation invariance and stabilizes the input to the fully connected layers. The dense latent bottleneck further compresses the 256-dimensional pooled representation into a 64-dimensional latent vector, enforcing a compact and physically meaningful encoding of the spectral response. This dimensionality reduction mitigates the "one-to-many" ambiguity of the inverse problem by forcing the network to capture only salient features that correlate with realizable MLMs geometries. Finally, the decoder, implemented as a deep fully-connected network, reconstructs the 12 geometric parameters from the latent vector, completing the inverse mapping from desired spectrum to structure (Figure 1c). The model is implemented in PyTorch and trained using the Adam optimizer with a learning rate of 0.001 and Mean Squared Error (MSE) loss between predicted and target geometric parameters. Training runs for 1000 epochs with a batch size of 64 on an NVIDIA GPU workstation. Table I summarizes the architectural layers and corresponding tensor shapes.

Electrodynamics physics of nonlinear SHG are incorporated into the design loop. For each predicted MLMs design, the linear reflection spectrum is first verified via FDTD to ensure that the dual resonances at the fundamental and second harmonic wavelengths match the target spectrum within a small tolerance. Subsequently, the nonlinear SHG response is calculated using Equations (1)–(7). The resulting SHG enhancement factor *EF* are compared for each design. Although the neural network is trained purely on linear spectra and geometric parameters, the nonlinear figures-of-merit are used to select and benchmark high-performance designs and to guide the creation of augmented training datasets around promising regions of the design space. The quality factors $Q_{\omega}$ and $Q_{2\omega}$ of the fundamental and second harmonic resonances, respectively, are obtained from Lorentzian fits to the simulated reflection spectra. The effective modal volumes $V_{\text{eff}}(\omega)$ and $V_{\text{eff}}(2\omega)$ are defined analogously to the linear case as

$$V_{\text{eff}}(\omega) = \frac{\int_V \varepsilon(\omega,r)|E(\omega,r)|^2 dV}{\max_r[\varepsilon(\omega,r)|E(\omega,r)|^2]} \tag{10}$$

and similarly for $V_{\text{eff}}(2\omega)$. These parameters allow us to relate the SHG enhancement to classical cavity figures-of-merit and to benchmark our designs against other nonlinear resonant structures in the literature

**TABLE I**: NanoPhotoNet-PINL model layers with different layers parameters

| Layer type | Input shape | Output shape | Parameters |
|---|---|---|---|
| Conv1D (1→32, kernel=9, stride=2) | (Batch, 1, 1000) | (Batch, 32, 500) | 320 |
| Conv1D (32→64, kernel=7, stride=2) | (Batch, 32, 500) | (Batch, 64, 250) | 14,400 |
| Conv1D (64→128, kernel=5, stride=2) | (Batch, 64, 250) | (Batch, 128, 125) | 41,088 |
| Conv1D (128→256, kernel=5, stride=2) | (Batch, 128, 125) | (Batch, 256, 63) | 163,840 |
| Conv1D (256→256, kernel=3, stride=2) | (Batch, 256, 63) | (Batch, 256, 32) | 196,864 |
| AdaptiveAvgPool1D | (Batch, 256, 32) | (Batch, 256, 1) | 0 |
| Squeeze | (Batch, 256, 1) | (Batch, 256) | 0 |
| Linear (256→128) | (Batch, 256) | (Batch, 128) | 32,896 |
| Linear (128→64) | (Batch, 128) | (Batch, 64) | 8,256 |
| Linear (64→64) | (Batch, 64) | (Batch, 64) | 4,160 |
| Linear (64→128) | (Batch, 64) | (Batch, 128) | 8,320 |
| Linear (128→256) | (Batch, 128) | (Batch, 256) | 33,024 |
| Linear (256→12) | (Batch, 256) | (Batch, 12) | 3,084 |

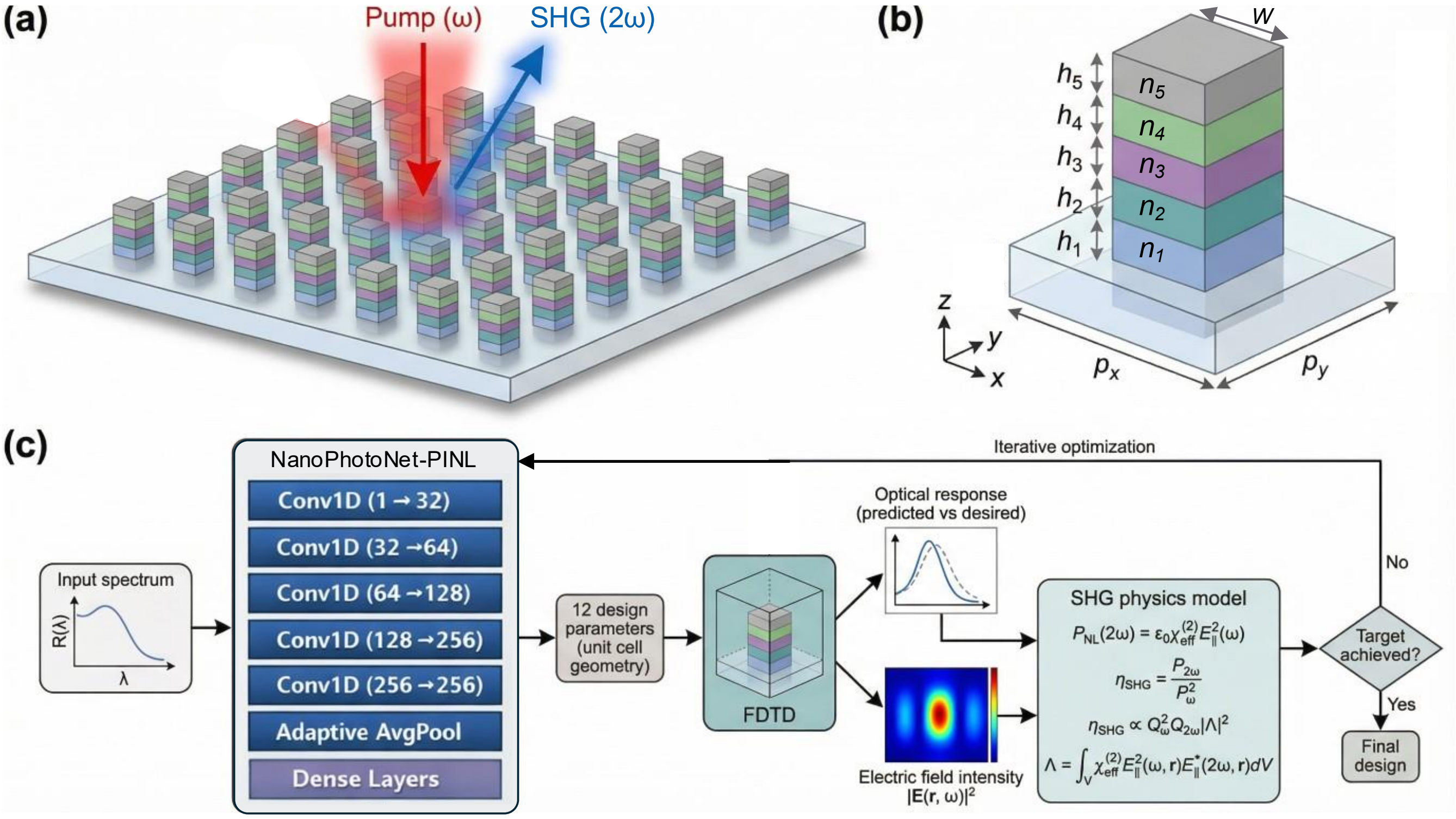

**FIG. 1.** Dual-resonance multi-layer metasurfaces using physics-informed AI-assisted inverse design for enhanced SHG. (a) Schematic of a dual-resonance MLMs cavity integrating a 3R-$MoS_2$ sheet at an internal interface, with an incident pump at frequency $\omega$ and radiated second harmonic at $2\omega$. (b) A representative MLMs unit cell with symmetric period ($P_x = P_y$), showing nanopillar width ($w$), layer heights ($h_i$), material refractive index ($n_i$), and 3R-$MoS_2$ placement in the center. (c) NanoPhotoNet-PINL architecture, consisting of a 1D convolutional encoder, adaptive average pooling, latent bottleneck, and dense decoder mapping target dual-resonant spectra to 12 MLMs design parameters. The physical integration to optimize the maximum mode overlapping and SHG conversion efficiency.

## 3. Results and Discussion

Figure 2 illustrates the progression of both training and validation losses for the NanoPhotoNet-PINL framework as a function of training epochs. A clear convergence trend is observed, where the training loss decreases monotonically during the early stages of optimization, reflecting the model's ability to effectively capture the highly nonlinear relationship between the input spectral features and the twelve target geometric parameters of the MLMs. The training loss rapidly approaches convergence and eventually saturates at approximately 1.2%. Simultaneously, the validation loss, used to evaluate the model's predictive capability on unseen MLM configurations, follows a similar declining trajectory, reaching a stable minimum of around 5.3%. The existence of a consistent gap between the training and validation losses can be attributed to the intrinsic complexity of the inverse design problem, particularly the ill-posed "one-to-many" nature of electromagnetic inverse scattering, where multiple structural configurations can produce nearly indistinguishable optical responses. Despite this challenge, the absence of divergence between the two loss curves and their stable asymptotic behavior indicate that the model maintains strong generalization performance without evidence of severe overfitting. This is further supported by the high prediction accuracy, with an efficiency exceeding 99.2%, confirming the reliability of the proposed framework for the inverse design and synthesis of novel MLMs geometries.

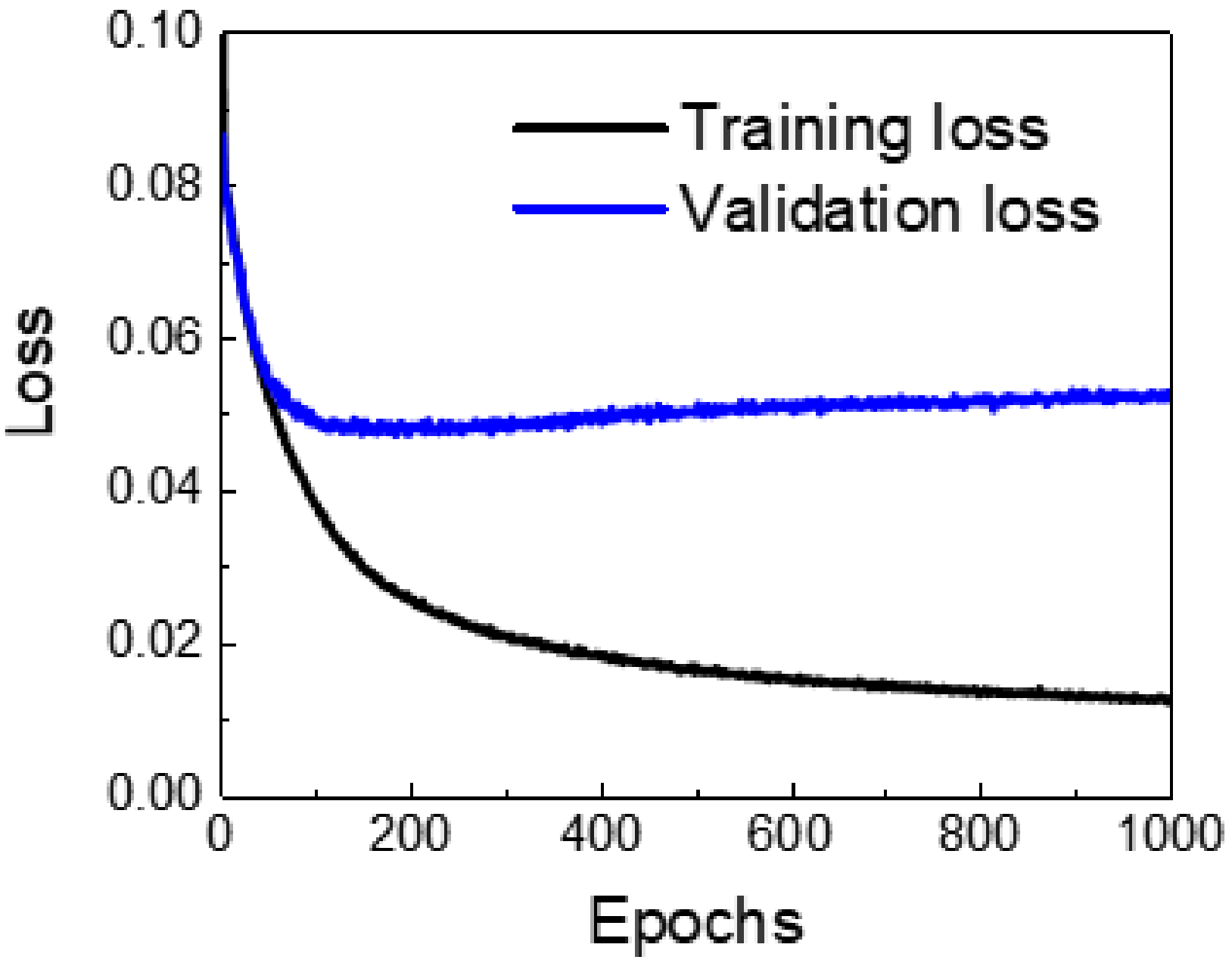

**FIG. 2.** Loss curves versus epochs of NanoPhotoNet-PINL model during training and validation.

### Section 3.1: On demand inverse design of single resonance MLMs cavity

To realize efficient SHG enhancement in nanostructured 3R-$MoS_2$ integrated within MLMs, it is critical to engineer resonant cavities that simultaneously support *Q*-factors and strong electromagnetic field confinement at both the pump and SH wavelengths. This requires systematic and iterative optimization of both geometric parameters and material compositions to meet stringent spectral and modal requirements. Following the completion of model training, the NanoPhotoNet-PINL framework was employed to inversely predict the optimal MLMs configurations corresponding to user-defined reflection spectra. Specifically, the model outputs twelve structural and material parameters, which are subsequently validated through FDTD simulations to ensure consistency between the predicted and actual optical responses, including both spectral characteristics and spatial field distributions. These validated spectrums and fields are then incorporated into a physics-based SHG model to compute the generated SH power and conversion efficiency. A performance checkpoint is applied to assess whether the predicted structure satisfies the desired criteria, such as resonance alignment, dual-resonance conditions, or reflection/transmission characteristics. If the target specifications are not achieved, the optimization loop is repeated until convergence is obtained (Figure 1c). To rigorously evaluate the inverse design capability of NanoPhotoNet-PINL, the model was tested on deterministic single-resonance targets spanning a broad spectral range from 400 nm to 800 nm (Figure 3). In the near-ultraviolet regime, the model successfully reconstructed a target resonance at 390 nm, yielding a structure with a *Q*-factor of 123. This performance was achieved through a Fabry–Perot-type cavity configuration based on an $Al_2O_3$–$Si_3N_4$–ZnO multilayer stack, leveraging the relatively low intrinsic optical losses of $Si_3N_4$ and ZnO in this spectral region to enhance resonance confinement. In the blue–green spectral range, for a target resonance near 500 nm, the model predicted a configuration with a *Q*-factor of 106. Here, the interplay between $Si_3N_4$ and ZnO was strategically utilized, where the moderate optical losses of ZnO contributed to controlled linewidth broadening and spectral shaping of the resonance. A key challenge in the inverse design process lies in incorporating the 3R-$MoS_2$ layer within the optimized MLMs structure, particularly due to its relatively high absorption at shorter wavelengths. Nevertheless, 3R-$MoS_2$ remains highly advantageous owing to its large refractive index, strong second-order nonlinear susceptibility, and comparatively lower losses at longer wavelengths (>700 nm). Accordingly, for a target resonance at 700 nm, the model identified an optimal structure incorporating 3R-$MoS_2$, achieving a *Q*-factor of 32 by appropriately scaling critical geometrical parameters such as periodicity and feature size. Extending further into the near-infrared regime, a target resonance at 800 nm resulted in a predicted structure with a *Q*-factor of 45, again utilizing 3R-$MoS_2$ as a key functional layer. The predicted reflection spectra shown in Figure 3a were systematically validated against FDTD simulations, demonstrating excellent agreement with an overall prediction accuracy exceeding 99.2%. Considering the multilayer nature of the structures, the total SH response arises from contributions across all constituent layers. Therefore, to quantify the practical SHG performance, the SH enhancement was evaluated for the complete optimized MLMs configurations using the nonlinear model described by Equations (1)–(7). Resulting SH emission spectrums are presented alongside the corresponding reflection responses within each subfigure. To benchmark performance, a comparative analysis was conducted against a reference unpatterned thin-film (TF) structure comprising identical material stacks and thicknesses as the optimized MLMs. This comparison reveals that the engineered MLMs cavities provide substantial SHG enhancement, reaching up to 123-fold improvement in the ultraviolet region using low-loss materials and approximately 111-fold in the visible range using 3R-$MoS_2$ relative to the TF baseline. Further improvement in SHG efficiency can be achieved through precise spectral alignment of the pump wavelength and, more importantly, by engineering dual-resonant conditions that simultaneously enhance both the fundamental and SH frequencies. Such dual-resonance designs are expected to significantly strengthen light–matter interactions, thereby maximizing nonlinear conversion efficiency.

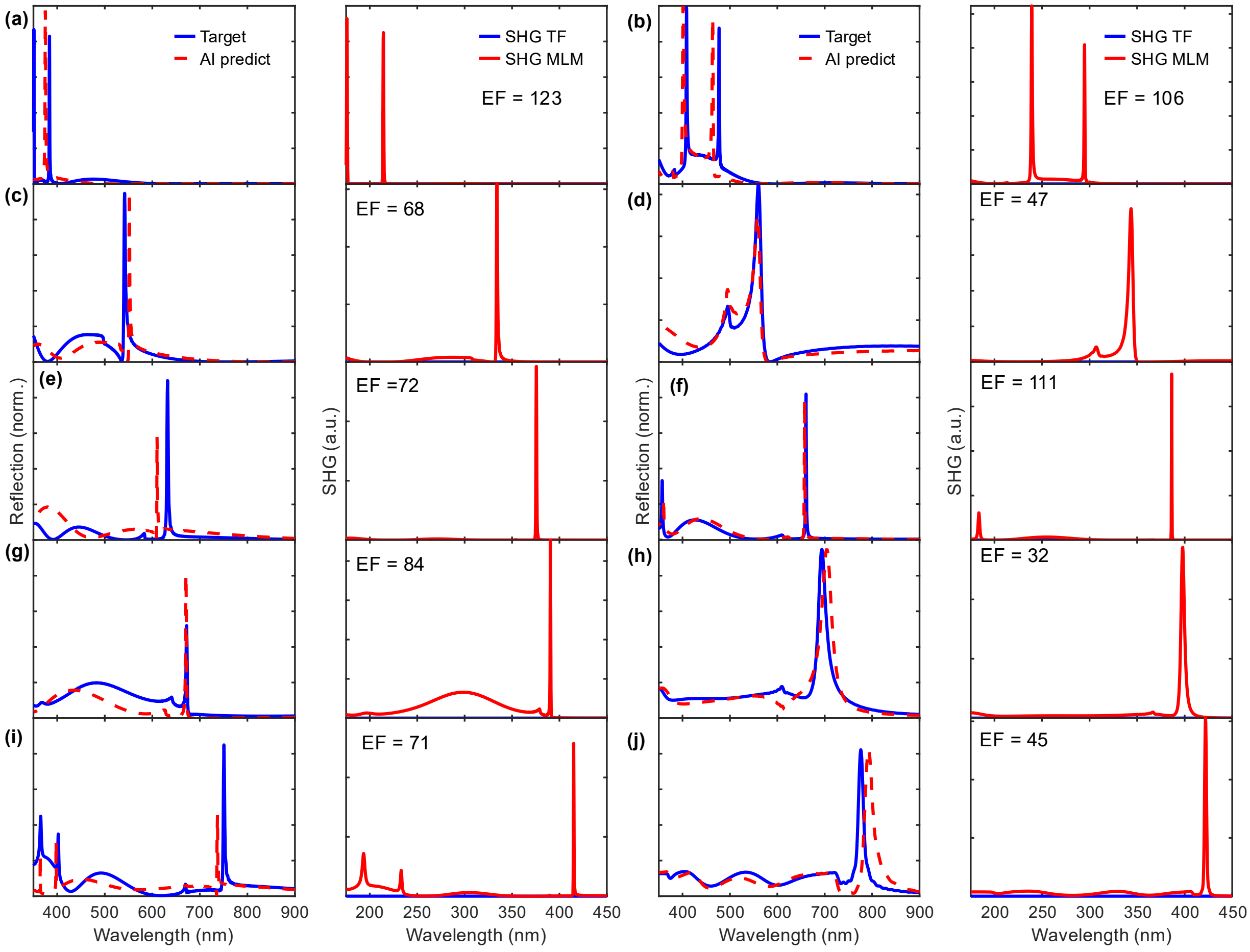

**FIG. 3.** On-demand inverse design of single-resonance cavities utilizing the NanoPhotoNet-PINL framework. For each designated target, the left sub-panel compares the desired reflection spectrum for a single-resonance cavity (blue curve) with the predicted reflection spectrum of MLMs generated by the NanoPhotoNet-PINL model (red curve). The corresponding right sub-panel illustrates the SHG response of the optimized MLMs (red curve) relative to an unpatterned thin film (blue curve), alongside the calculated SHG enhancement factor. The inverse-designed single-resonance MLM cavities are presented for target wavelengths of (a) 390 nm, (b) 400 nm, (c) 500 nm, (d) 540 nm, (e) 560 nm, (f) 650 nm, (g) 680 nm, (h) 690 nm, (i) 700 nm, (j) 750 nm, and (k) 800 nm.

Figure 4 presents a comprehensive comparison between the target and inverse-designed electric field distributions for single-resonance MLMs structures across a broad spectral wavelength range as illustrated in Figure 3. For each design, the left panel defines the desired (True) spatial electric field profile, specifically engineered to maximize field localization within MLMs, thereby enhancing nonlinear light–matter interactions. The central panel illustrates the corresponding field distribution obtained from the NanoPhotoNet-PINL-predicted MLMs configuration, while the right panel quantifies the spatial deviation between the target and predicted fields. Across all investigated wavelengths, the predicted MLMs designs consistently exhibit strong electromagnetic field confinement within the multilayer stack, with pronounced localization in selected layers, particularly in proximity to the center region. This behavior highlights the capability of the inverse design framework to engineer spatially tailored optical modes, offering a powerful route to manipulate light–matter interactions in complex nanophotonic architectures. Such controlled confinement is not only essential for enhancing nonlinear optical processes, but also represents an emerging paradigm for designing functional photonic devices with spatially engineered responses. It is important to note that the reference (True) field distributions are obtained from full-wave FDTD simulations based on the ground-truth set of twelve design parameters, whereas the predicted fields are generated using the corresponding parameters inferred by the NanoPhotoNet-PINL model. Despite this indirect reconstruction, the predicted field profiles demonstrate excellent agreement with the target distributions across all cases. A clear scaling trend is also observed as a function of wavelength. Structures designed for shorter wavelengths (e.g., 390–500 nm) exhibit reduced periodicity and thinner layer thicknesses, consistent with the need to support resonances at higher frequencies. In contrast, MLMs targeting longer wavelengths (e.g., 700–800 nm) feature larger periods and increased thicknesses to accommodate the extended optical path lengths required for resonance formation. The spatial error maps further confirm the robustness of the model, with only minor discrepancies localized primarily at material interfaces and edge regions. These deviations can be attributed to small mismatches between the predicted and true geometric parameters, as well as the presence of finite material losses. Nevertheless, the overall low error magnitude underscores the high predictive fidelity of the model. The demonstrated ability to accurately reproduce and control electric field distributions opens new opportunities for tailoring optical confinement in advanced applications, including light-emitting metasurfaces, nonlinear frequency conversion platforms, and phase-engineered flat optics.[28, 29]

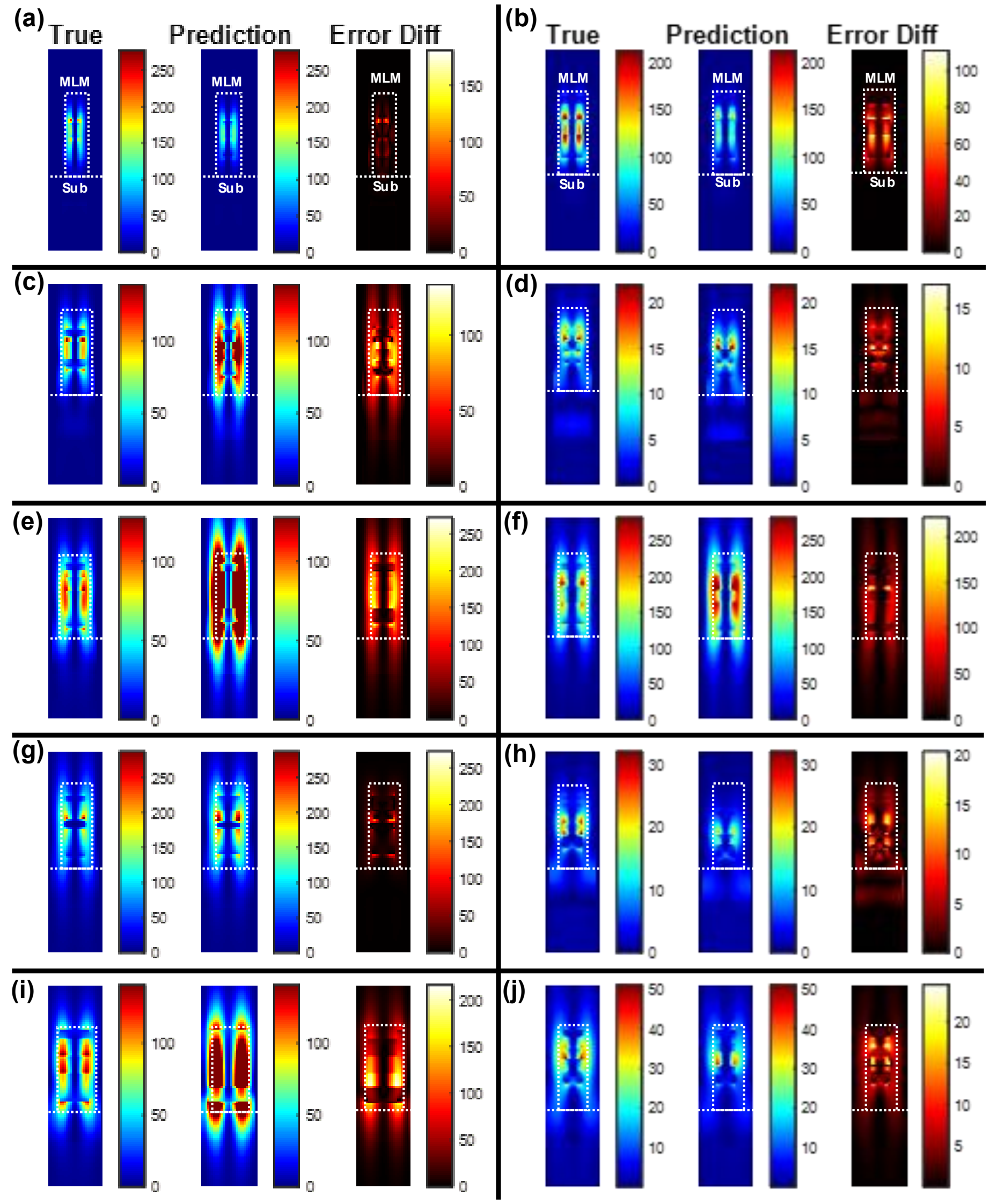

**Fig. 4.** Desired spatial distributions of the electric field profiles for the inverse-designed single-resonance MLMs. For each designated target, the left panel illustrates the desired (True) electric field profile, emphasizing the strong optical field confinement within and immediately surrounding the 3R-$MoS_2$ layer. The center panel displays the corresponding predicted electric field profile using NanoPhotoNet-PINL for the optimized design. The right panel maps the spatial error difference between the desired and predicted field distributions. Profiles are presented for target resonance wavelengths of (a) 390 nm, (b) 400 nm, (c) 500 nm, (d) 540 nm, (e) 560 nm, (f) 650 nm, (g) 680 nm, (h) 690 nm, (i) 700 nm, (j) 750 nm, and (k) 800 nm.

## Section 3.2: Inverse design of on demand dual resonances MLM cavity

The ultimate capability of the NanoPhotoNet-PINL framework is demonstrated through the deterministic synthesis of MLMs cavities engineered for phase-matched dual-resonance operation. These structures are specifically designed to simultaneously enhance light–matter interaction at both the fundamental (pump) wavelength and the corresponding second-harmonic (SH) wavelength of 3R-$MoS_2$. A series of optimized dual-resonance configurations were inversely designed within the spectral range of 700–900 nm, which coincides with the peak nonlinear response of 3R-$MoS_2$. For the first design shown in Figure 5a, targeting a pump wavelength of 700 nm, the framework generated an MLMs structure exhibiting dual reflection resonances at both the pump and SH wavelengths, with Q-factors of approximately 53 and 5, respectively. This performance was achieved using a ZnO–$Al_2O_3$–3R-$MoS_2$–$TiO_2$ multilayer stack, where the high refractive index of 3R-$MoS_2$ and the low intrinsic optical losses of ZnO, $Al_2O_3$, and $TiO_2$ collectively contribute to efficient field confinement across both spectral regions. The resulting SHG enhancement factor, relative to a thin-film (TF) reference, reached approximately 265. Achieving optimal phase-matching conditions, particularly when balancing strong confinement at shorter wavelengths with material absorption constraints, typically requires extensive iterative optimization using conventional numerical methods. Such approaches often involve numerous simulation cycles and significant computational cost. In contrast, NanoPhotoNet-PINL enables rapid exploration of the design space, performing large-scale iterative optimization at speeds exceeding full-wave FDTD simulations by more than four orders of magnitude. As shown in Figure 5b, the model successfully identified a phase-matched configuration at 716 nm, achieving a substantial SHG enhancement factor of 6030. Similarly, Figures 5c–5e present additional optimized designs with enhancement factors of 4860 at 745 nm, 8103 at a longer wavelength, and 5320, respectively. Notably, the highest enhancement (Figure 5d) is attributed to the formation of two vertically coupled Fabry–Perot resonant cavities, independently tuned to the pump and SH wavelengths, thereby maximizing both excitation efficiency and nonlinear emission. As a result, these results demonstrate the high accuracy, efficiency, and deterministic nature of the NanoPhotoNet-PINL framework in designing complex MLMs architectures with precise phase-matching and dual-resonance characteristics. The significant improvements in SHG efficiency validate the potential of this approach for advancing integrated nonlinear nanophotonic devices, particularly in the development of compact and high-performance nanoscale light sources.

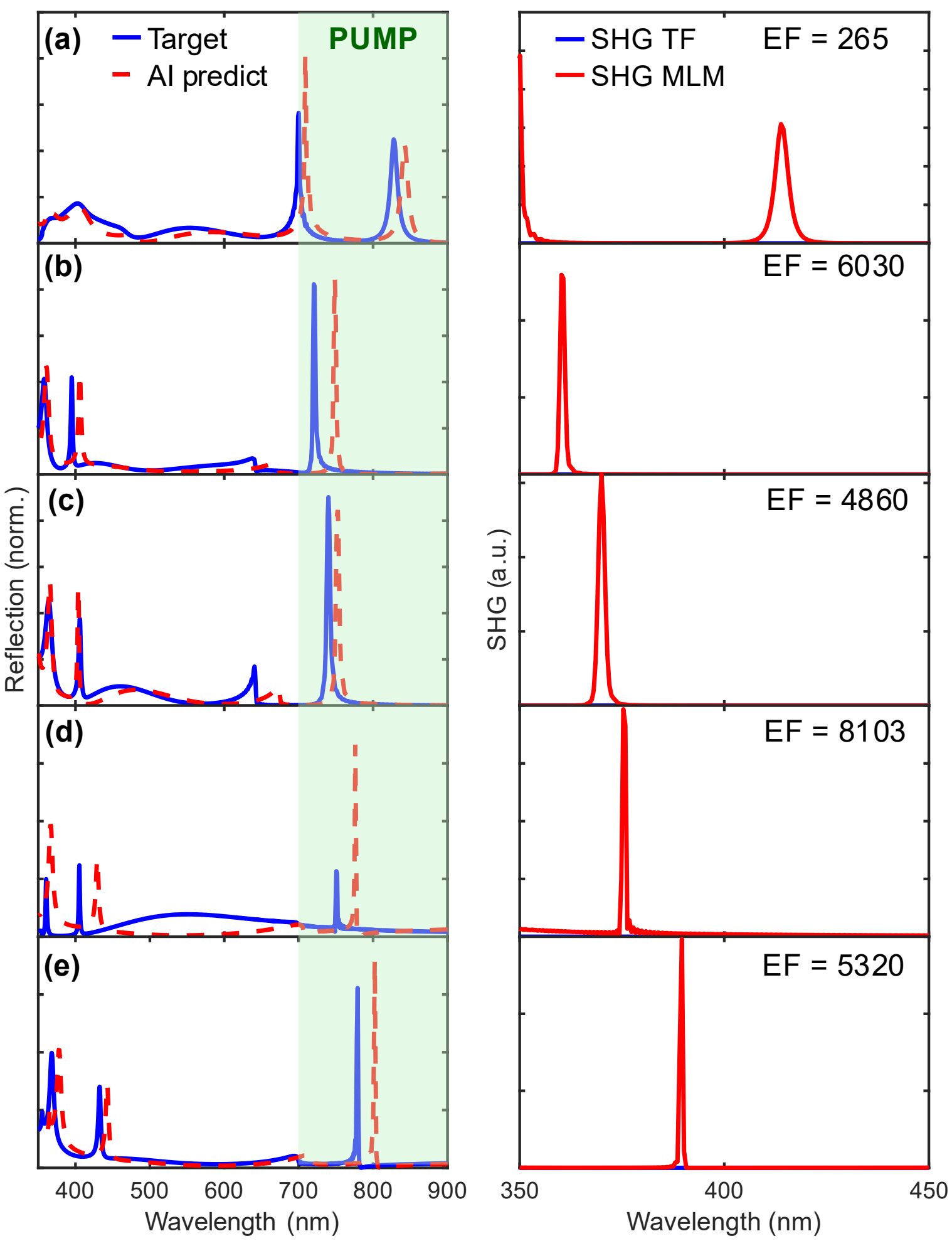

**Fig. 5.** On-demand inverse design of modal phase-matched dual-resonance cavities utilizing the NanoPhotoNet-PINL framework. For each designated target, the left sub-panel compares the desired reflection spectrum for a dual-resonance cavity at the pump and corresponding SH wavelengths (blue curve) with the predicted reflection spectrum of the multilayer metasurfaces (MLMs) generated by the NanoPhotoNet-PINL model (red curve). The corresponding right sub-panel illustrates the SHG response of the optimized MLMs (red curve) relative to an unpatterned thin film (blue curve), alongside the calculated SHG enhancement factor. The inverse-designed phase-matched dual-resonance MLM cavities are presented for target wavelengths of (a) 700 nm, (b) 716 nm, (c) 745 nm, (d) 780 nm, and (e) 810 nm.

Figure 6 provides a detailed assessment of the spatial electric field distributions for the inverse-designed phase-matched dual-resonance MLMs configurations at both the pump and corresponding second-harmonic (SH) wavelengths. For each design case, the left panel defines the target field profile, intentionally engineered to achieve strong localization of the electromagnetic field within the multilayer stack at pump wavelength, particularly in the vicinity of the embedded 3R-$MoS_2$ layer. The central panel presents the field distribution obtained from the NanoPhotoNet-PINL-predicted structure, while the right panel quantifies the spatial deviation between the target and reconstructed fields. Across all investigated designs, the predicted dual-resonance MLMs consistently demonstrate pronounced field confinement not only within the overall multilayer architecture but also selectively within specific functional layers, most notably around the nonlinear 3R-$MoS_2$ region. This simultaneous confinement at both the fundamental and SH wavelengths highlights the capability of the framework to engineer coupled resonant modes, thereby enabling enhanced and controllable light–matter interaction. Such dual-wavelength field localization represents a powerful and emerging strategy for designing advanced nanophotonic systems with tailored nonlinear and quantum optical responses. For the SH response, the electric field distributions are not directly simulated but are instead computed using a nonlinear optics equations Equations (1)–(7) that incorporates the intrinsic second-order susceptibility and material properties of the MLMs stack. This approach ensures a physically consistent evaluation of the nonlinear emission profiles. The spatial error maps reveal minimal discrepancies between the target and predicted fields across all cases, with deviations primarily localized at material interfaces and boundary regions. These small differences arise from minor variations between the predicted and true design parameters, as well as the influence of finite material losses. Nevertheless, the low overall error magnitude confirms the strong predictive capability of the model in capturing both linear and nonlinear field distributions. Importantly, the demonstrated accuracy in reproducing complex, dual-resonant field profiles provides a foundation for extending this approach to other applications requiring precise field engineering. In particular, the ability to tailor electromagnetic confinement at multiple frequencies is highly relevant for quantum photonic platforms, including the development of efficient single-photon sources and other nonlinear light-emitting nanostructures.[29]

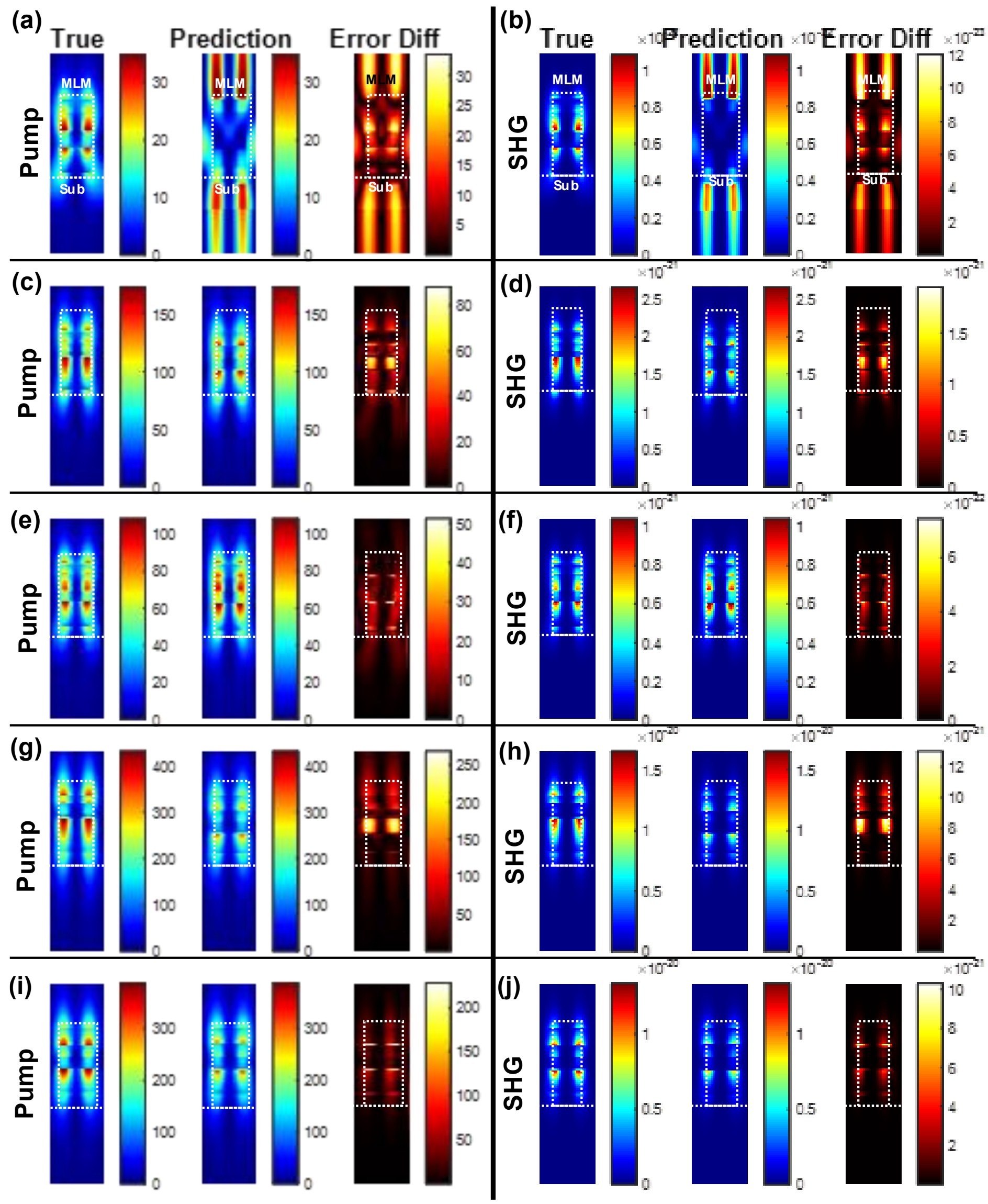

**Fig. 6.** Spatial distributions of the electric field profiles for the inverse-designed phase-matched dual-resonance MLMs. For each sub-figure, the left panel illustrates the desired electric field profile, emphasizing strong optical field confinement within and immediately surrounding the 3R-$MoS_2$ layer at the specified wavelength. The center panel displays the corresponding predicted electric field profile for the optimized design. The right panel maps the spatial error difference between the desired and predicted field distributions. Profiles are presented at both the pump and corresponding SH wavelengths for each target design: (a) pump and (b) SH profiles for the 700 nm target; (c) pump and (d) SH profiles for the 716 nm target; (e) pump and (f) SH profiles for the 745 nm target; (g) pump and (h) SH profiles for the 780 nm target; and (i) pump and (j) SH profiles for the 810 nm target.

Following the realization of spectrally aligned dual-resonance MLM cavities incorporating 3R-$MoS_2$ and operating near the peak of its second-order nonlinear susceptibility, the resulting enhancement in SH output was quantitatively evaluated as a function of the pump power and benchmarked against reference TF 3R-$MoS_2$ structures reported in the literature. Achieving high SH power is essential for the development of compact and efficient nanoscale light sources for next-generation photonic systems. The SH response was computed using the nonlinear formalism described in Equations (1)–(7), with appropriate calibration to experimentally reported TF 3R-$MoS_2$ data,[57] enabling a physically consistent estimation of SH power. For the unstructured TF configuration, the maximum SH photon count was on the order of $10^5$. In contrast, the inverse-designed dual-resonance MLMs structures exhibited a substantial increase, reaching SH photon counts on the order of $10^9$. This corresponds to an enhancement factor approaching ~8,103-fold, as illustrated in Figure 7a. To further contextualize this performance, a comparison was conducted with state-of-the-art nanophotonic platforms designed to enhance SHG in 3R-$MoS_2$ without the use of AI-assisted inverse design. Previous studies have reported enhancement factors ranging from ~$10^3$ to $2.2\times10^3$ for bound-state-in-the-continuum (BIC) metasurfaces (Figure 7b),[58] and approximately 140-fold enhancement using anapole-resonant metastructures.[59] In integrated photonic waveguide platforms incorporating 3R-$MoS_2$, the enhancement is typically limited to ~5-fold due to relatively low $Q$-factor resonances associated with grating-based designs.[60] In comparison, the proposed MLMs dual-resonance framework significantly exceeds these benchmarks. The enhanced performance arises from a synergistic optimization strategy that simultaneously maximizes pump field confinement and strengthens nonlinear light–matter interaction through spectrally matched resonances at both the fundamental and SH frequencies. As summarized in Table II, the optimized MLM designs achieve SHG enhancement factors of up to $8.1\times10^3$, with $Q$-factors reaching 108 at the pump wavelength and 111 at the SH wavelength. These values substantially outperform those reported in prior studies. Overall, these results highlight the effectiveness of integrating multilayer structural flexibility with AI-driven inverse design methodologies, enabling unprecedented performance improvements in nonlinear nanophotonic devices.

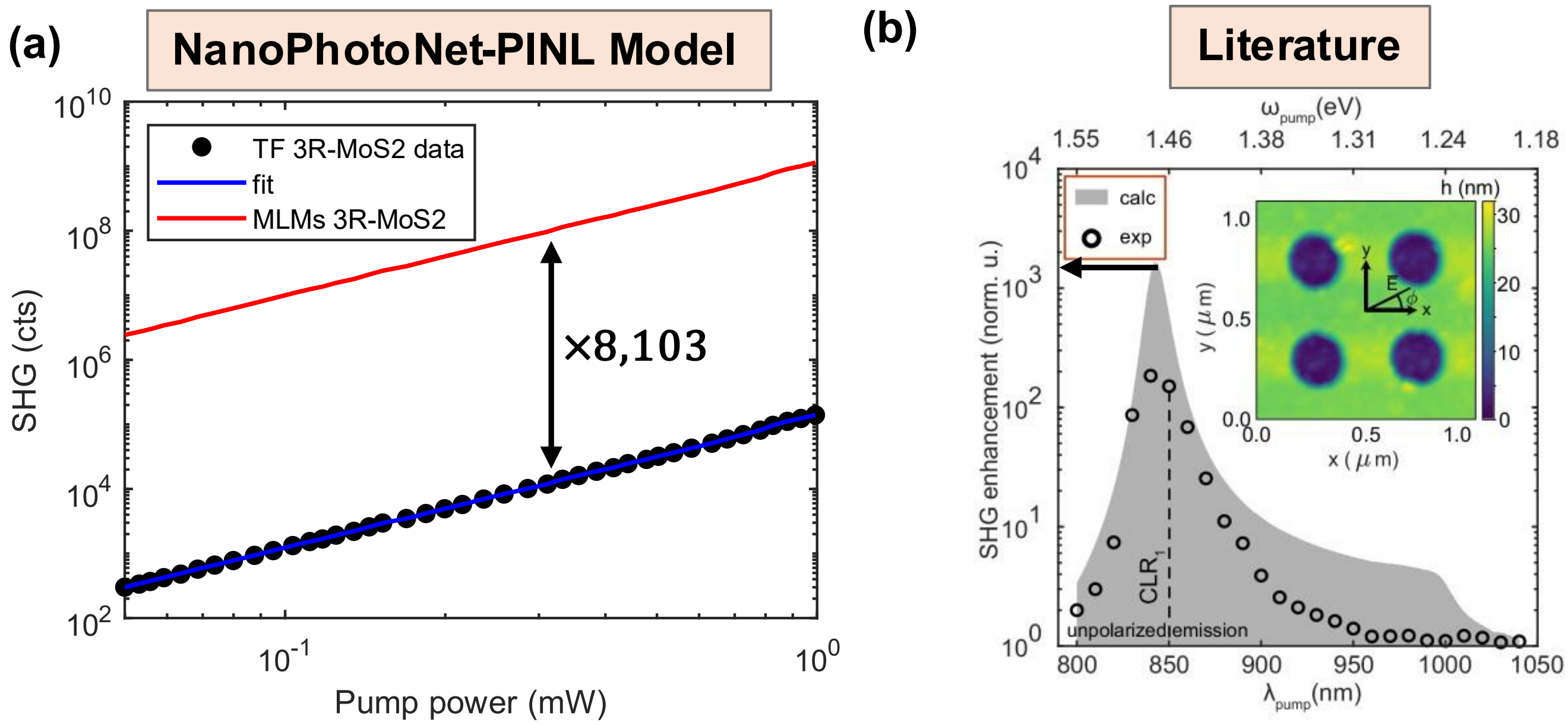

**Fig. 7.** Benchmarking SHG performance of the NanoPhotoNet-PINL designs against state-of-the-art 3R-$MoS_2$-based metasurfaces supporting BICs resonance. (a) Pump power dependence of the enhanced SHG for the inverse-designed modal phase matched MLMs, compared with the experimentally measured SHG of an unpatterned 3R-$MoS_2$ thin film reported in the literature.[57] (b) Comparison of the wavelength-dependent SHG enhancement against a 3R-$MoS_2$ metasurface supporting BIC resonances optimized via conventional numerical methods.[58] Reproduced with permission from Springer, Copyright 2025.

**Table II:** Benchmark phase match MLMs performance with literature SHG 3R-$MoS_2$ nanocavities

| Metasurface design | AI-design method | 3R-$MoS_2$ thickness (nm) | Pump Q-factor | SHG Q-factor | SHG λ (nm) | SHG enhancement (a.u.) | Reference |
|---|---|---|---|---|---|---|---|
| Single layer MS – BIC | No | 20 ~ 25 | 100 | - | 400 ~ 450 | $10^3$ | 58 |
| Single layer MS – Anapole | No | 160 | 57 ~ 79 | - | 825 | 140 | 59 |
| Single layer MS – BIC | No | 213 | 100 ~ 150 | - | 820 | $2.2\times10^3$ | 61 |
| Integrated Wave Guide | No | 20 ~ 50 | 10 | - | 445 | 5 | 60 |
| **Multilayer MS (MLM)** | **NanoPhotoNet-PINL** | **25 ~ 81** | **51 ~ 108** | **5~111** | **350 ~ 416** | **265 ~ $8.1\times10^3$** | **This work** |

The predictive performance of the NanoPhotoNet-PINL framework was systematically benchmarked against prior studies on artificial intelligence–driven inverse design in nanophotonics (Table III). Existing approaches, particularly those applied to relatively simple and low-dimensional geometries such as nanoholes or cylindrical meta-atoms, typically report prediction accuracies in the range of approximately 84%–95%. In contrast, the proposed framework operates within a substantially more complex design space involving MLMs architectures with higher dimensionality and stronger parameter interdependence, yet achieves a superior prediction accuracy of 99.2%. This improvement can be attributed to the effectiveness of the 1D convolutional encoder coupled with a latent bottleneck representation. Beyond predictive accuracy, NanoPhotoNet-PINL demonstrates a significant computational advantage over conventional full-wave numerical methods. Specifically, the framework provides an acceleration of approximately $5 \times 10^4$ compared to FDTD simulations executed under identical computational conditions. This substantial speed-up effectively eliminates the need for extensive iterative simulations, which would otherwise require thousands of computational hours for comprehensive parameter sweeps and optimization. Consequently, the proposed approach enables a transition from computationally prohibitive, trial-and-error-based design methodologies to a rapid and deterministic inverse design paradigm. This capability is particularly critical for scalable device engineering and high-throughput optimization in emerging applications, including integrated quantum photonics,[30] nonlinear optical platforms,[36] and industrial metasurface manufacturing.[42]

**Table III:** Performance comparison between literature inverse AI models and NanoPhotoNet-PINL.

| AI architecture | Accuracy (%) | Reference |
|---|---|---|
| MetasurfaceViT | 85 | 62 |
| MGDNN | 83.9 | 63 |
| MetaFAP | 92.7 | 64 |
| GAF–CNN–LSTM | 95 | 65 |
| **NanioPhotoNet-PINL** | **99.2** | **This work** |

## 4. Conclusion

In conclusion, we have developed NanoPhotoNet-PINL, a physics informed AI-driven inverse-design framework for nonlinear MLMs that achieves high-efficiency second harmonic generation in 3R-$MoS_2$ integrated into dual-resonant MLMs cavities. By combining a 1D CNN-based encoder for spectral feature extraction, GAP layer, and a DNNs decoder for structural parameter prediction, the model accurately maps desired dual-resonant reflection spectra to multi-layer geometries with prediction accuracy exceeding

~99.2%. Unlike conventional trial-and-error or forward-only optimization, NanoPhotoNet-PINL efficiently navigates the high-dimensional design space of MLMs and yields structures that simultaneously optimize the fundamental and second harmonic resonances, the spatial field overlap, and the radiation pattern. By embedding nonlinear electrodynamics into the design loop, we quantify SHG enhancement via overlap integrals and cavity figures-of-merit, demonstrating that dual-resonant MLMs can achieve SHG amplification factors up to 8103-fold compared to a planar 3R-$MoS_2$ bare flake under identical pumping conditions. These enhancements surpass or match those of state-of-the-art anapole and quasi-BIC nonlinear metasurfaces while providing greater flexibility in engineering multiple resonances and far-field responses. Furthermore, the AI-enabled framework delivers several orders of magnitude reduction in design time relative to full FDTD-based optimization, thereby enabling rapid prototyping of nonlinear metasurfaces tailored to specific application requirements. Looking ahead, the NanoPhotoNet-PINL paradigm can be extended to other second-order and third-order processes, including sum-frequency generation, difference-frequency generation, and Kerr nonlinearities in emerging 2D materials and wide-bandgap semiconductors. Incorporating additional degrees of freedom such as phase-gradient engineering, tunable materials, and active modulation could enable dynamically reconfigurable nonlinear metasurface platforms. Such developments will further advance integrated nonlinear nanophotonics and support next-generation applications in on-chip frequency combs, quantum light sources, and ultrafast all-optical information processing.

## Credit Authorship Contribution Statement

**Omar A. M. Abdelraouf:** Project administration, Methodology, Conceptualization, Investigation, Data curation, Validation, Supervision, Resources, Visualization, Writing – original draft & review.

## Declaration of Competing Interest

The authors declare no competing financial interests or personal competing interest.

## Data Availability

NanoPhotoNet-PINL code and raw data are available on request.